# Structural and spin-glass properties of single crystal $J_{\text{eff}}$ = ½ pyrochlore antiferromagnet NaCdCo$_2$F$_7$ : correlating $T_\text{f}$ with magnetic-bond-disorder

A. Kancko,[1] G. Giester,[2] and R. H. Colman [1*]

[1] *Charles University, Faculty of Mathematics and Physics, Department of Condensed Matter Physics, Ke Karlovu 5, 121 16 Prague 2, Czech Republic*
[2] *Institut für Mineralogie und Kristallographie, Universtät Wien, Austria*

*Corresponding author:ross.colman@mag.mff.cuni.cz*

## Abstract:

Weak bond disorder disrupts the expected spin-liquid ground-state of the ideal $S = 1/2$ Heisenberg pyrochlore antiferromagnet. Here we introduce a single crystal study of the structural and magnetic properties of the bond-disordered pyrochlore NaCdCo$_2$F$_7$. The magnetic susceptibility appears isotropic, with a large negative Curie-Weiss temperature ($\theta_{\text{CW}}$ = -108(1) K), however no magnetic order is observed on cooling until a spin-glass transition at $T_\text{f}$ = 4.0 K. AC-susceptibility measurements show a frequency-dependent shift of the associated cusp in χ' at $T_\text{f}$, that can be fitted well by the empirical Vogel-Fulcher law. The magnetic moment of $\mu_{\text{eff}}$ = 5.4(1) μ$_B$/Co$^{2+}$ indicates a significant orbital contribution and heat capacity measurements show that down to 1.8 K, well below $T_\text{f}$, only $S_{\text{mag}}$ ~2/3 Rln(2) of the magnetic entropy is recovered, suggestive of residual continued dynamics. Structural and magnetism comparisons are made with the other known members of the Na$A$''Co$_2$F$_7$ family ($A$'' = Ca$^{2+}$, Sr$^{2+}$), confirming the expected relationship between spin-glass freezing temperature, and extent of magnetic bond disorder brought about by the size mismatch between $A$-site ions.

**Keywords:** Pyrochlore, Cobalt, Frustrated magnetism, Spin-glass, Structural disorder

## Introduction

Properties of $S$ = ½ Heisenberg pyrochlore antiferromagnet remains hotly contentious as it is a notoriously difficult system to model theoretically using current methods. The ground-state predictions are strongly method-dependent and range from dimer singlet phases [1–3], spin-liquids with short-range correlations [4,5], as well as chiral spin liquids [6].

To complement and confirm the theories, model materials must be found and studied, as close as possible to the ideal case. Pyrochlore lattices of magnetic ions exist in a number of real materials, most commonly in both the spinel, AB$_2$O$_4$, and pyrochlore, A$_2$B$_2$O$_7$, oxides [7–11]. Structural stability and charge balancing constraints of the pyrochlore oxides, mean that the magnetic ions that are most commonly hosted are rare-earth $A^{3+}$ ions. Several rare-earth pyrochlore oxides have been proposed as having spin-liquid ground-states [12–14], although the contracted nature of the 4$f$ orbitals means that magnetic exchange energies are typically low (typically 0.1 - 1 meV), and very low temperatures are required to study the spin-liquid-state properties.

A relatively less studied family is that of the pyrochlore fluorides $A$'$A$''$M_2$F$_7$, where: $A$' = a monovalent cation, $A$'' is a divalent cation, and $M$ is a divalent 3$d$ transition metal ion. The magnetic properties of several members of this group (with $A$' = Na$^+$, $A$'' = Ca$^{2+}$ or Sr$^{2+}$, and $M$ = Co$^{2+}$, Ni$^{2+}$, Fe$^{2+}$ and Mn$^{2+}$) have already been reported, with only the Co and Ni members extensively studied [15–18]. The more spatially extensive 3$d$ orbitals result in greater orbital overlap and significantly stronger magnetic exchange interactions (typically 10 – 100 meV), evidenced by the relatively large Curie-Weiss temperatures so far observed ($|\theta_{\text{CW}}|$ ~120-140 K).

The noteworthy deviation from ideality in these fluoride pyrochlores is the necessity for mixed ion occupancy on the $A$-site ($A$'$^+$ and $A$''$^{2+}$), due to the charge balancing requirement for an average $A^{1.5+}$ oxidation state. No structural ordering of these $A$-site ions has been found, which leads to small local-structure deviations from the average structure because of ion size mismatches between $A$' and $A$''. These local structural deviations transcribe onto the magnetic interactions too, introducing magnetic exchange disorder that eventually results in spin-glass freezing rather than the fully-dynamic spin-liquid state predicted by some models in an ideal S=1/2 Heisenberg pyrochlore antiferromagnet.

Despite the apparent spin-glass transition at 4 K, the $S = 1$ NaCaNi$_2$F$_7$ shows persistent spin dynamics down below 80 mK and a continuum of excitations observed by inelastic neutron scattering suggests a large manifold of low-energy states [19,20]. A similar picture is found in both $J_{eff}$ = ½ NaCaCo$_2$F$_7$ and NaSrCo$_2$F$_7$, where a short-range-ordered state is selected from a continuous manifold of low-energy states and frozen in at $T_f$ = 2.4 K and 3.0 K, respectively [21,22]. Comparison of Monte Carlo simulations with NMR data confirms the importance of quantum fluctuations and single-ion anisotropy in driving the low-temperature properties of these systems [23].

In this paper our single crystal study introduces a new, isostructural and isoelectronic member of this $J_{eff}$ = ½ pyrochlore family: NaCdCo$_2$F$_7$. We present the synthesis and structural analysis, comparing with the previously studied Co$^{2+}$ pyrochlore antiferromagnets. Magnetic and thermodynamic investigations characterise the spin-freezing transition, and we consider the structure-property relationships that drive the spin-freezing transition temperature across this series.

## Experimental methods

Single crystals of NaCdCo$_2$F$_7$ were grown using a laser floating zone furnace (Crystal Systems Corp. FZ-LD-5-200W-II-VPO-PC). Dry elemental fluorides were weighed out and ground in a glove-box due to their hygroscopic nature. The mixture was then filled into a hollow graphite tubular crucible and melted in the laser furnace under an 8 bar atmosphere of argon, at ~1000 °C (using infrared pyrometry and assuming an emissivity of 0.85 for graphite), to form a polycrystalline precursor rod. After cooling, the rod was extracted from the graphite crucible and re-mounted within the furnace for standard floating-zone growth using platinum wire. The growth was again performed under a dynamic flow of argon atmosphere (0.25 l/min) at high pressure (8 bar) to minimize evaporation, with melting point ~750 °C (assuming an emissivity of 0.72). The red oligocrystalline ingot was broken into smaller pieces, many of which were single grain fragments. A small selection of the grains was crushed into a fine powder and analysed by Powder X-Ray Diffraction (PXRD) using a Bruker D8 Advance diffractometer with Cu K$\alpha$ radiation ($\lambda$ = 1.5418 Å), and a structural refinement was performed using Topas Academic V6 [24]. Neighbouring pieces were used for both single crystal property measurements and single crystal X-ray diffraction analysis. Single crystal diffraction was performed at 200 K using a Bruker Kappa X8 APEX diffractometer equipped with a CCD detector and Incoatec Microfocus Source IμS (30 W, multilayer mirror, Mo$K\alpha$). All atomic structure figures were prepared using Vesta 3 [25]. X-ray fluorescence (XRF) spectroscopy was performed using an EDAX AMETEK ORBIS-PC spectrometer equipped with a Rh anode tube ($E_{K\alpha}$ = 20.216 keV) and Apollo XRF ML-50 EDS detector with polycapillary focusing optics (~184 ×69 μm$^2$ spot size). Orienting of the crystals for magnetic and physical property measurements was carried out using a PhotoScience Laue diffractometer. Temperature and field dependent DC susceptibility measurements on the oriented crystals were performed in a Quantum Design Magnetic Property Measurement System (model MPMS-XL 7T), using the reciprocating sample option (RSO). Temperature and frequency dependent AC susceptibility measurements were made using the ACMS II option in a Quantum Design Physical Property Measurement System (PPMS). Heat capacity was measured via the heat relaxation method also in the PPMS, with the addition of a $^3$He low-temperature insert, using a non-magnetic sapphire stage and Apiezon-N grease for mounting the crystal. For subtraction of the lattice contribution to specific heat, a sample of the non-magnetic analogue NaCdZn$_2$F$_7$ was also prepared using the synthesis techniques described above for NaCdCo$_2$F$_7$.

## Results

### Structure

The single crystal reflections could be indexed by space-group $Fd$-$3m$ with cell parameter $a$ = 10.3488(5) Å. The solved structure, summarised in Table 1, is isostructural to that of NaSrCo$_2$F$_7$ [16], NaCaCo$_2$F$_7$ [15], and the other recently studied pyrochlore fluorides shown in figure 1(a) [17,18]. As with these previous examples, no superstructure reflections were seen, indicating no ordering of Na and Cd on the pyrochlore (16$d$) $A$-site. The Cd:Co ratio was confirmed on several of the single crystal pieces by XRF spectroscopy to be 1:2.0(1), and was constrained as this ideal stoichiometry for the reported structure below. However, the single crystal refinement showed a preference for a slightly increased electron density on the Co (16$c$) site. We attempted to model this by joint Co:Cd occupation which lead to a ratio of 0.959(2):0.041(2), respectively, a final composition of NaCd$_{1.08}$Co$_{1.92}$F$_7$ with a drop in $R$-factor from 1.55 to 1.19 %, compared to the ideal stoichiometry case. To further confirm the single crystal solution is representative of the bulk, powder X-ray diffraction was collected on a specimen of powder after crushing several grains. Figure 1(c). shows the results of Rietveld refinement using the ideal stoichiometry pyrochlore structure solution.

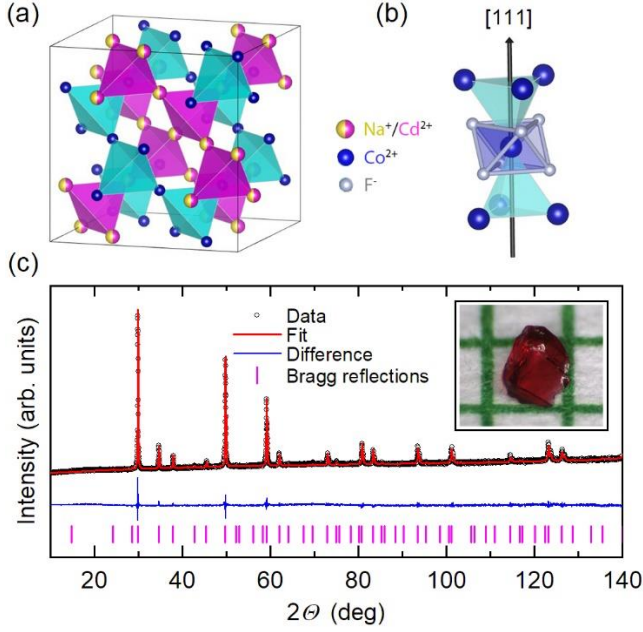

FIG. 1. (a) Crystal structure of $NaCdCo_2F_7$, highlighting the pyrochlore lattice of $S_{eff} = ½$ $Co^{2+}$ ions, with $F^-$ ions left off for clarity, and the intertwined pyrochlore lattice of the mixed-occupancy $Na^+/Cd^{2+}$ A-site. (b) The local octahedral coordination environment of the $CoF_6$ unit, superimposed on vertex-sharing tetrahedral magnetic exchange lattice. (c) Powder diffraction data of crushed $NaCdCo_2F_7$ crystal, collected at 300 K. Rietveld refinement using the pyrochlore structure finds $a = 10.3636(2)$ Å with a goodness-of-fit, $\chi^2 = 0.42$.

Table 1. Single crystal structure solution.

| | Space group: $Fd\bar{3}m$ (227, origin 2) | | $a = 10.3488(5)$ Å | | | $V = 1108.3(3)$ Å$^3$ | | | $Z = 8$ | | |
|---|---|---|---|---|---|---|---|---|---|---|---|
| | Radiation: Mo Kα | | $T = 200$ K | | | Reflections collected/unique: 17497/179 | | | Parameters: 11 | | |
| | Goodness of fit: 1.274 | | Final R indices: $R = 1.55$ %, $wR = 3.95$ % | | | Largest difference: Peaks Holes | | | 0.535 e Å$^{-3}$ -0.542 e Å$^{-3}$ | | |
| Atom | Site | x/a | y/a | z/a | Occ. | $U_{11}$ /(Å$^2$ x10$^{-3}$) | $U_{22}$ | $U_{33}$ | $U_{23}$ | $U_{13}$ | $U_{12}$ |
| Na | 16d | 0.5 | 0.5 | 0.5 | 0.5 | 16.02(14) | 16.02(14) | 16.02(14) | -2.93(6) | -2.93(6) | -2.93(6) |
| Cd | 16d | 0.5 | 0.5 | 0.5 | 0.5 | 16.02(14) | 16.02(14) | 16.02(14) | -2.93(6) | -2.93(6) | -2.93(6) |
| Co | 16c | 0 | 0 | 0 | 1 | 8.31(13) | 8.31(13) | 8.31(13) | -0.33(5) | -0.33(5) | -0.33(5) |
| F(1) | 8b | 0.375 | 0.375 | 0.375 | 1 | 17.0(5) | 17.0(5) | 17.0(5) | 0 | 0 | 0 |
| F(2) | 48f | 0.33415(15) | 0.125 | 0.125 | 1 | 25.1(7) | 20.2(4) | 20.2(4) | 10.5(5) | 0 | 0 |

## Magnetic susceptibility

Temperature-dependent DC magnetic susceptibility measurements (determined as $\chi = M/H$ in a constant field $H = 2000$ Oe), with the field applied along the [100], [110] and [111] crystallographic directions of $NaCdCo_2F_7$, are shown in Figure 2(a). In all calculations, the ideal stoichiometry, $NaCdCo_2F_7$, was used. The data in all three high symmetry directions appear Curie-Weiss-like and overlay, suggesting an isotropic magnetic behaviour $NaCdCo_2F_7$. The inverse susceptibility $\chi^{-1}(T)$ data with H ∥ [111] (Figure 2(b)) were used for a Curie-Weiss fit in the temperature range of 100 - 350 K, in the standard form of $\chi = C/(T-\Theta_{CW})$, where $C = N_A\mu_{eff}^2/3k_B$ is the Curie constant, $N_A$ is Avogadro's number, $\mu_{eff}$ is the effective magnetic moment, $k_B$ is Boltzmann's constant and $\Theta_{CW}$ is the Curie-Weiss temperature. The bottom right inset shows a linear dependence of magnetisation vs. field (at $T = 2$ K) up to $\mu_0H = 7$ T, justifying our calculation of susceptibility as $M/H$ at 2000 Oe. The Curie-Weiss fit of the temperature-dependent susceptibility data yields a large negative Curie-Weiss temperature (a mean-field measure of the interaction strength) of approximately $\Theta_{CW} \approx -108(1)$ K, confirming dominant antiferromagnetic interactions in $NaCdCo_2F_7$. The effective moment per cobalt was extracted as $\mu_{eff} \approx 5.4(1)$ $\mu_B$, which is significantly larger than the expected $S = 3/2$ $Co^{2+}$ spin-only value (3.87 $\mu_B$), and is approaching the $J = 9/2$ total angular momentum value (6.63 $\mu_B$), as was seen previously in the other isostructural analogues. [15,16]

A cusp at $T = 4.0$ K seen in the inset of Figure 2(a), concomitant with bifurcation of the ZFC and FC data indicates a magnetic transition with history dependence, suggestive of either magnetic ordering or spin freezing into a disordered state – a spin glass.

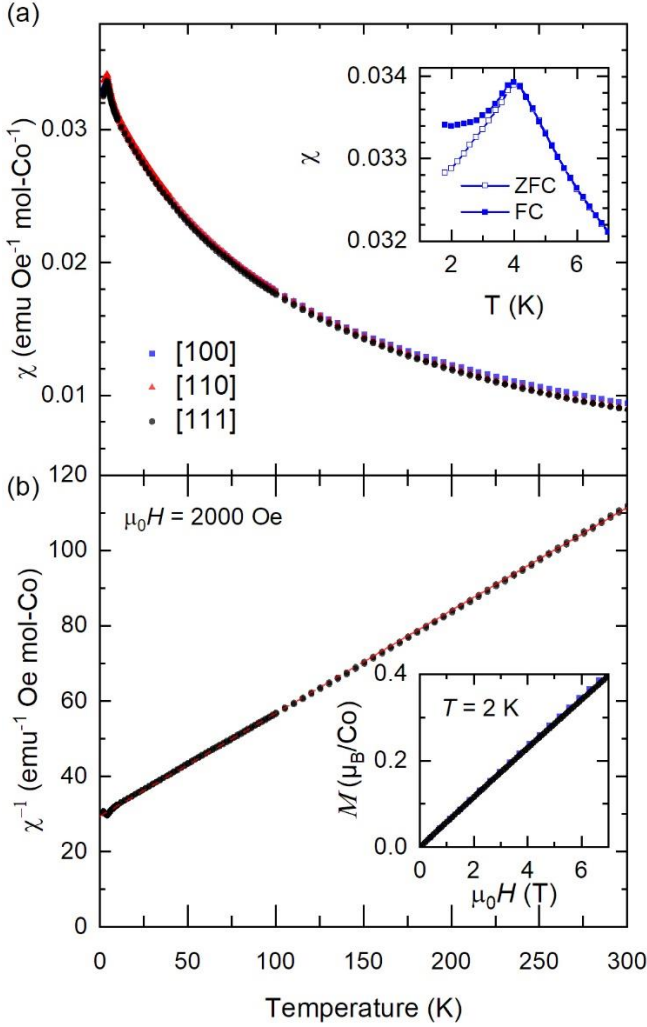

FIG. 2(a) Temperature-dependent DC-magnetic-susceptibility for NaCdCo$_2$F$_7$ along significant crystallographic directions at $\mu_0 H$ = 2000 Oe. The inset highlights a low-temperature cusp. (b) Inverse susceptibility for $H \parallel$ [111], with Curie-Weiss fit (red line). The inset shows $M(H)$ at 2 K, indicating a linear dependence of the magnetic response of all crystallographic directions up to $\mu_0 H$ = 7 T.

To further probe the observed 4 K cusp in DC measurements, AC susceptibility measurements were performed. The low temperature region shows again a cusp in χ', with a weak but well defined frequency-dependent shift, indicative of a spin-glass ground-state (figure 3 inset). The extracted dimensionless Mydosh parameter follows the relative shift of $T_f$ per decade of frequency, $K = \Delta T_f / (T_f \Delta \log \nu) = 0.010(1)$ [26]. This value is smaller than those expected for superparamagnetic or cluster-glass states, is in line with those seen in the isostructural analogues [15,16], and is typical of other insulating spin-glass systems [27,28]. Additional inferences about the properties of the spin-glass are typically extracted by fitting the frequency-dependent shift to the Vogel-Fulcher law, which takes into account the interactions of clusters of spins during the dynamic freezing process [26]. The Vogel-Fulcher law expects a frequency-dependence of $T_f = T_0 - \frac{E_a}{k_B} \frac{1}{\ln(\tau_0 \nu)}$, where $T_0$ is the ideal glass temperature, $E_a$ is the activation energy of the transition and $\tau_0$ is the intrinsic relaxation time. Since we are fitting a restricted frequency range, free fitting of all parameters is not stable and instead $\tau_0$ is typically fixed to a reasonable value of $10^{-12}$ s. These parameters can be extracted from a fit of $T_f$ vs $-1/\ln(\tau_0 \nu)$, resulting in $E_a = 6.6(6) \cdot 10^{-4}$ eV and $T_0 = 3.7(1)$ K, shown in figure 3.

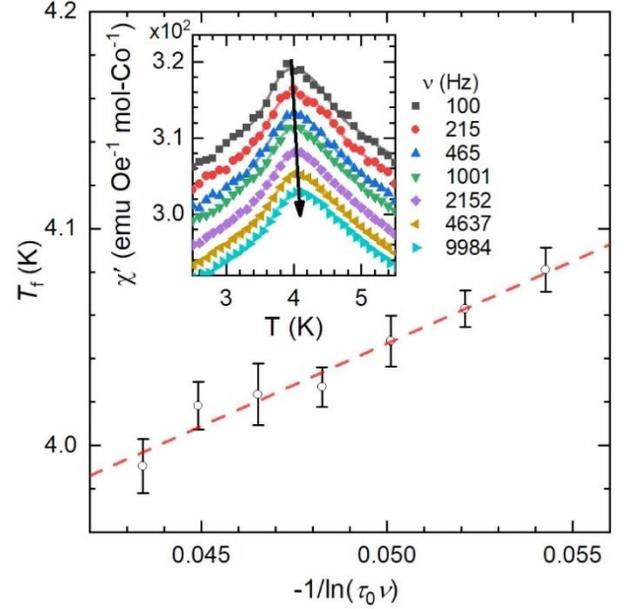

FIG 3. AC-susceptibility of NaCdCo$_2$F$_7$, showing the frequency dependence of a peak in χ' as a function of both temperature and frequency, $\nu$, in the inset. With the frequency dependence of the freezing temperature, $T_f$, fitted according to the Vogel-Fulcher law in the main panel.

## Specific heat

The temperature dependence of the specific heat is shown in Figure 4(a). The lattice contribution was subtracted using an isostructural non-magnetic analogue NaCdZn$_2$F$_7$ [29]. With no contribution to the total heat capacity from conduction electrons due to the insulating nature of NaCdCo$_2$F$_7$, the resulting magnetic contribution to the specific heat, $C_{mag}$, can be approximated as

$$C_{mag} = C_p - C_{lattice} \approx C_{p[NaCdCo_2F_7]} - C_{p[NaCdZn_2F_7]}.$$

A significant magnetic contribution can be seen in Figure 4(b) at low temperatures, starting on cooling below ~40 K with a maximum in $C_{mag}/T$ vs. $T$ (Figure 4 (a) inset) at 4.0 K. The high temperature (>100 K) apparent additional contribution to $C_{mag}/T$ is likely due to transitions to the first Kramers' doublet excited state, found to be 25.8(5) meV in isostructural NaCaCo$_2$F$_7$ [22].

Below $T_f$ the magnetic contribution to specific heat shows an exponential decay (figure 4(b)). The $T < T_f$ data was fitted to a simple exponential function, $C_{mag} = AT^\alpha$, resulting in values of $A = 0.115(10)$ $J$ mol-Co$^{-1}$ and exponent $\alpha = 1.744(7)$. A linear dependence of $C_{mag}(T)$ is expected below $T_f$ according to most canonical spin-glass models and has been observed in the metallic spin glasses [26,30]. Exponential behaviour has, however, been seen previously in several geometrically frustrated spin-glass systems, including NaCaNi$_2$F$_7$ [19,31,32].

The magnetic entropy, $S_{mag}(T)$, is shown in Figure 4(c), and is consistent with the assignment of an $S_{eff} = 1/2$ magnetic ground-state, approaching but noticeably short of even a value of Rln(2). This is far below the value expected for an orbitally quenched spin-only $S = 3/2$ state value of $S_{mag} = $ Rln(4). The suppression of the peak with increasing applied field, seen in the inset of Figure 4(a), is typical of antiferromagnetic spin-glass systems [30].

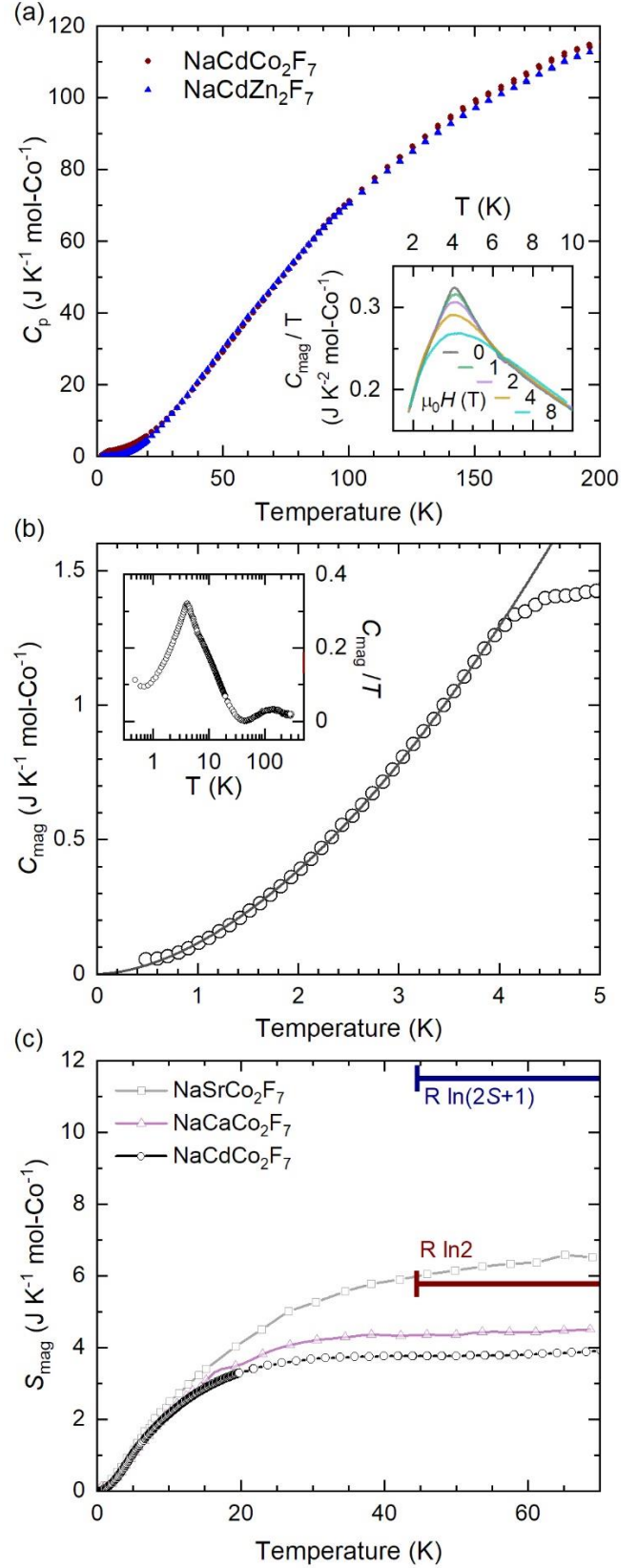

FIG 4(a). Temperature dependent heat capacity of NaCdCo$_2$F$_7$ and non-magnetic analogue NaCdZn$_2$F$_7$ in zero

field. Inset shows a zoom of the low temperature region of the magnetic component, $C_{mag}/T$, for NaCdCo$_2$F$_7$ with multiple applied fields ($H \parallel [100]$). (b) shows the exponential decay of $C_{mag}$ below $T_f$, with line showing the fit to $C_{mag} \propto T^{\alpha}$, with $\alpha$ = 1.743(7). The inset shows the expanded temperature dependence of $C_{mag}/T$ on a logarithmic scale. (c) The change in magnetic entropy, $S_{mag}$, saturating at ~2/3 Rln(2), with comparisons to previously published data for related NaSrCo$_2$F$_7$ and NaCaCo$_2$F$_7$, extracted from [16] and [15], respectively.

## Discussion

First we must consider the structure of NaCdCo$_2$F$_7$, and compare its details with those of its isoelectronic analogues NaSrCo$_2$F$_7$ and NaCaCo$_2$F$_7$. The structure matches that of the other pyrochlore fluorides, with a lattice parameter slightly less than empirical predictions using the $A$ and $B$-site ion sizes ($a_{exp}$ = 10.3636(2) Å, $a_{pred}$ = 10.422 Å, for experimental and predicted, respectively) [33]. It is worth noting that several other Cd-containing pyrochlores were omitted during creation of the empirical model as they were also systematically lower than those predicted.

Full site disorder of the $A$-site is still maintained on mixing of $A$'' = Cd$^{2+}$ with $A$' = Na$^+$. Using the VIII coordination Shannon ionic radii for each ion [34], the ion size discrepancy, $\delta_{rA} = 2(r_{A'} - r_{A''})/(r_{A'} + r_{A''})$ is -6.6 %, 5.2 %, and 7.0 % for Na$^+$ with Sr$^{2+}$, Ca$^{2+}$ and Cd$^{2+}$, respectively. Whilst the $A$-site is not directly involved in magnetic exchange, variation of the $A$-site ions can introduce subtle changes to the structure that indirectly effect the magnetic properties discussed in the following paragraphs.

The pyrochlore structure has very limited allowed variation as only one crystallographic parameter is not constrained by symmetry – the F2 (48$f$) site $x$-coordinate, $x_{F2}$. All other atoms are fully constrained to high-symmetry positions. Two easy to describe extremes for the $x_{F2}$-coordinate exist, illustrated in Figure 5(a), resulting in a situation ranging from: an ideal octahedral coordination of the $BF_6$ group, with a highly distorted $AF_8$ polyhedron, at $x_{F2}$ = 0.3125 on the left; to a trigonally compressed $BF_6$ triangular antiprism, with an ideal cube $AF_8$ unit, at $x_{F2}$ = 0.375 on the right. In NaCdCo$_2$F$_7$ $x_{F2}$ = 0.33415(15), which results in a variation of the F-Co-F bond angles of $\pm\delta_{\text{F-Co-F}}$ = 8.12(5)° away from the ideal 90°, due to the trigonal compression along the [111] axis. The $x_{F2}$ coordinate, and subsequently the extent of trigonal compression of the CoF$_6$ octahedron, is seen to correlate with the average $A$-site ionic radii, showing increasing trigonal compression as the ionic radius decreases from $A$'' = Sr$^{2+}$, towards Cd$^{2+}$.

When a Co$^{2+}$ (3d$^7$) ion is in an octahedral crystal field, the degenerate ($^4$F) ground state is first split into $^4$A$_2$ (highest), $^4$T$_2$, and $^4$T$_1$ (lowest) terms. With a trigonal distortion and spin-orbit coupling, this $^4$T$_1$ term can further split into six Kramers doublets, with an effective $S_{eff}$ = 1/2 doublet as the ground state [35]. The low-magnetic entropy associated with the spin-freezing transition (<Rln(2)) seems to confirm this spin-orbit-coupled ground-state, in agreement with the previous spectroscopic studies of the isostructural family members [22].

Increasing the trigonal distortion of the Co$^{2+}$ octahedron has previously been shown to increase the $g$-factor anisotropy [36]. A similar situation has been proposed in the rare-earth iridate pyrochlores $A_2$Ir$_2$O$_7$, where *chemical pressure* through rare-earth substitution on the $A$-site leads to increased tetragonal distortion that increases spin-anisotropy and promotes a magnetically ordered state [37–39]. Although the Heisenberg $S$ = 1/2 pyrochlore ground-state is predicted to be a quantum spin-liquid [40], increasing single-ion anisotropy to either XY-planar or uniaxial Ising has the potential to drive the system to order [41,42]. In our case of NaCdCo$_2$F$_7$, the extent of magnetic anisotropy is not yet established, but our low-field magnetisation measurements, where χ(T) of [100], [110], and [111] overlay, suggest isotropic magnetism. Similarly, both low-field and high-field (up to $\mu_0 H$ = 60 T) magnetisation measurements on NaCaCo$_2$F$_7$ also overlay perfectly, suggesting low-to-zero magnetocrystalline anisotropy, confirmed with electron spin resonance spectroscopy measurements that give an isotropic $g$-factor of ~2 [43]. These result are, however, inconsistent with inelastic neutron scattering data that are interpreted with a model showing short-range XY-like correlations with a g-tensor anisotropy of $g_{xy}/g_z$ ~ 3 in NaCaCo$_2$F$_7$ [21,22,44]. The increasing distortion of the CoF$_6$ octahedron, seen in Figure 5(a), suggests that NaCdCo$_2$F$_7$ is likely to have the largest anisotropy (most XY-like) of the so-far investigated members of this family. The presence of Cd in the compound significantly complicates the use of neutrons to investigate this however.

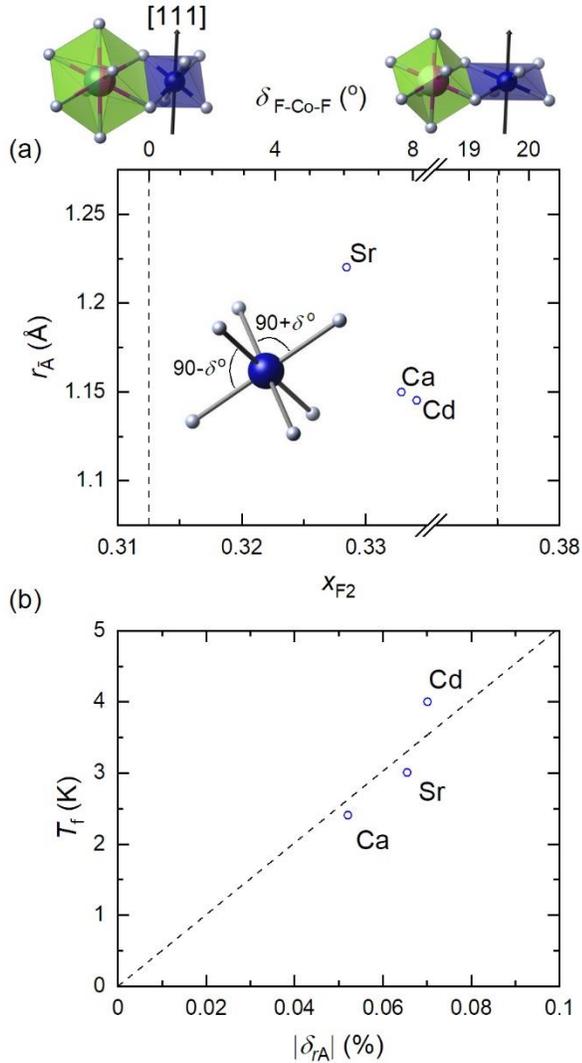

FIG. 5. Comparison of the level of distortion away from a regular octahedron of the $CoF_6$ triangular antiprisms in the Na$A$''$Co_2F_7$ series, with $A$'' = $Sr^{2+}$, $Ca^{2+}$ and $Cd^{2+}$. (b) Correlation between $A$-site ion size discrepancy, $|\delta_{rA}|$, and the spin-glass freezing temperature, $T_f$.

The inherent structural disorder due to mixed $A$-site occupancy in the $A'A''M_2F_7$ pyrochlores is held responsible for the spin-glass ground-state found in all members so far. Monte Carlo simulations of weak magnetic-exchange disorder, $\Delta$, in the pyrochlore Heisenberg antiferromagnet predict a spin-glass ground-state, with $T_f \propto \Delta$ when $\Delta \ll J$ [45,46]. The observed increase in $T_f$ across the series of $Ca^{2+}$, $Sr^{2+}$ to $Cd^{2+}$, from 2.4 K and 3.0 K to 4.0 K, where $A$-site ion-size mismatch, $|\delta_{rA}|$ increases from 5.2 % and 6.6 % to 7.0 %, respectively, supports this conclusion (Figure 5(b)). In this scenario we are of course considering the $A$-site ions to be purely spectators here, ignoring any potential additional electronic effects that may occur due to the presence of a filled 3$d$-shell in the case of $Cd^{2+}$ occupying the $A$-site.

The increasing $A$-site disorder is also correlated with both a drop in the Curie-Weiss antiferromagnetic coupling strength from $\theta_{CW}$ = -140 K in NaCaCoF$_7$ to -108 K in NaCdCoF$_7$; and the extracted effective moment from 6.1 to 5.4 $\mu_B$/Co, respectively, in the Curie-Weiss fits. The reduction in Curie-Weiss temperature in concert with the increase in spin-freezing temperature results in a reduction in the Ramirez frustration index, $f = |\theta_{CW}|/T_f$, from 58 where $A$'' = $Ca^{2+}$ to 27 for $Cd^{2+}$, although any 3-dimensional magnet with $f > 10$ can be considered highly geometrically-frustrated [47].

## Conclusions

We have reported the single-crystal growth, structural parameters, thermodynamic and magnetic properties measurements of a new pyrochlore $S_{eff}$ = ½ pyrochlore antiferromagnet, NaCdCo$_2$F$_7$. Despite the large negative Curie-Weiss temperature of -108 K, no sign of magnetic ordering occurs on cooling, until a spin-freezing transition is observed at $T_f$ = 4.0 K. The spin-glass nature of the transition is confirmed by a frequency-dependent shift in the $\chi$'-cusp. The transition is also observed in heat capacity measurements, with a characteristic broadening of the associated anomaly with increasing applied field.

Through comparison to the previously studied isostructural analogues, NaSrCo$_2$F$_7$ and NaCaCo$_2$F$_7$, the effects of subtle structural changes can be better understood. Increasing the size variance between randomly distributed $A$-site ions increases the magnetic exchange interaction disorder and pushes the system further from the expected spin-liquid ground-state in an ideal Heisenberg pyrochlore antiferromagnet. The increasing spin-freezing transition temperature is correlated with increasing $A$-site ion-size mismatch.

Unlike many 3$d$-transition metals, the magnetic moment in NaCdCo$_2$F$_7$ is significantly larger than the spin-only contribution so cannot be considered orbitally quenched. For $Co^{2+}$ in a trigonally distorted octahedral crystal field, spin-orbit coupling leaves an $S_{eff}$ = ½ Kramers doublet as the lowest energy state, supported by the low magnetic entropy of the spin-freezing transition ($S_{mag} \sim \frac{2}{3}$ Rln(2)). Such low recovered entropy suggests continued dynamics to lower temperatures, as was observed in NaCaNi$_2$F$_7$ [19].

The anisotropy of the magnetic exchange interaction in this family of materials is an ongoing problem with conflicting experimental interpretations. The expansion of the materials available for study, particularly in single crystal form, will provide important additional comparisons. Regrettably, the neutron absorption cross-section of natural Cd will prevent some of the most informative investigations such as direct determination of crystal field parameters [22,39], however numerous alternative spectroscopic techniques remain available. Understanding how much these materials deviate from an ideal $S = 1/2$ Heisenberg pyrochlore antiferromagnetic system will be an important consideration when comparing this materials' properties to the continuously developing theoretical predictions for the Quantum Pyrochlore Antiferromagnet, a notoriously difficult-to-model system.

## *Acknowledgements*

This work was supported by the Czech Science Foundation (project no. 19-21575Y). The preparation, characterization and measurement of bulk physical properties were performed in MGML (http://mgml.eu/), which was supported within the program of Czech Research Infrastructures (project no. LM2018096), as well as the OP VVV project MATFUN (number CZ.02.1.01/0.0/0.0/15_003/0000487). The authors additionally thank Gaël Bastien and Milan Klicpera for useful discussions.